\documentclass[12pt]{article}         
\usepackage{amsfonts}
\usepackage{amsmath}
\usepackage{latexsym}
\usepackage{color}
\usepackage{cite}
\usepackage{graphicx}[12pt]
\newcommand{\bm}[1]{\mbox{\boldmath$#1$}}

\DeclareMathOperator{\tg}{tg}

\newcommand{\bea}{\begin{eqnarray}}
\newcommand{\eea}{\end{eqnarray}}
\newcommand{\nn}{\nonumber}

\def\be{\begin{equation}}
\def\ee{\end{equation}}
\def\bea{\begin{eqnarray}}
\def\eea{\end{eqnarray}}
\def\nn{\nonumber}

\title{Effective potential energy for relativistic particles in the field of inclined rotating magnetized sphere}
\author{V. Epp$^{1,2}$\thanks{epp@tspu.edu.ru} and M. A. Masterova$^1$\\
\it{$^1$Tomsk State Pedagogical University, 634061 Tomsk, Russia}\\
\it{$^2$Tomsk State University,  634050 Tomsk, Russia}}
\date{ }
\begin{document}
\maketitle
\begin{abstract}
The dynamics of a charged relativistic particle in electromagnetic field of a rotating magnetized celestial body with the magnetic axis inclined to the axis of rotation is studied. The covariant Lagrangian function in the rotating reference frame is found.  Effective potential energy is defined on the base of the first integral of motion. The structure of the equipotential surfaces for a relativistic  charged particle is studied and depicted for different values of the dipole moment. It is shown that there are trapping regions for the particles of definite energies.

\vspace{2pt}
\end{abstract}

\section{Introduction}\label{sec1}
Motion of the charged particles in the field of a magnetized rotating celestial body is of large practical significance for astrophysics.
For example,  a charged particle in the Earth magnetic field is moving within the closed  regions which are  named radiation belts \cite{Alfven,Holmes}.
The trajectories of a charged particle in the dipolar magnetic field where studied in the papers \cite{Stormer07, Stormer55, DeVogelaere} and \cite{Dragt}.

  More complicated case is the case when direction of the magnetic moment differs from  direction of axis of rotation. In this case  an  electric field is induced inside and outside of the body. The neutron stars and pulsars are examples of such objects. The first model of electric field
 which is generated  in the neighbourhood of a neutron star was developed by Deutsch \cite{Deutsch}. Some other models were suggested and studied by several authors \cite{Michel}. Most of these models are based on assumption that the neutron star is a conducting sphere.  Electromagnetic field in this case differs essentially from the pure
 dipole field.
  Magnetic field of such objects  in good approximation can be described as the field of an inclined rotating magnetized sphere or ''oblique rotator"  \cite{Babcock}. Theoretical study of the field of an oblique rotator has a long history. We discuss it in more details in Section \ref{sec2}.

Allowed and forbidden regions of the motion of charged particles in such field was studied by Katsiaris and Psillakis \cite{Katsiaris}.
Dynamics of  a charged particle  near the force-free surface of a rotating magnetized sphere was explored in \cite{Thielheim_1989, Istomin}.
Some issues of charged particle dynamics within the electromagetic vacuum fields of an inclined rotator have already been discussed in the papers \cite{Ferrari_1975, Finkbeiner87}.
 
 Effective potential energy for a non-relativistic particle in the field of inclined rotating dipole was investigated in details in our recent paper \cite{Epp(2013)}, further referred to as Paper I. The calculations were made for the near region, i.e. for distances much less then the radius of the light cylinder. In the present paper we study the structure of the  effective potential energy for a relativistic particle in the field of inclined rotating magnetized sphere at the distances up to the light cylinder.

In Section 2 we show that the field of such sphere calculated by different authors under different assumption coincides with the field of rotating magnetic dipole if the sphere is uniformly magnetized and rotates with a non-relativistic speed.
In Section \ref{sec3} we present the relativistic Lagrange function for a charged particle in the arbitrary,  uniformly rotating electromagnetic field. The integral of motion of such Lagrange system is calculated. Existence of the integral of motion gives the possibility to introduce an effective potential energy which allows studying some general features of the particle motion without solving the equations of motion. Section \ref{sec4} represents analysis of the main properties of the effective potential energy. The equipotential surfaces are obtained by numerical calculations and demonstrated in pictures of  Section \ref{sec5}. Section \ref{sec6} contains discussion of the results and our conclusions.
%%%%%%%%%%%%%%%%%%%%%%%%%%%%%%%%%
%%
\section{The electromagnetic field of a rotating magnetized sphere}\label{sec2}
%%
%%%%%%%%%%%%%%%%%%%%%%%%%%%%%%%%%%
In this section we analyse the field of rotating magnetized sphere. There are different ways of modelling  the field of an oblique rotator.

Deutsch describes a non-relativistic  rotating magnetized star as a
perfectly conducting sphere in rigid rotation in vacuo \cite{Deutsch}. In order to introduce a relativistic model of the field source  Belinsky et al. \cite{Belinsky_Ruffini, Belinsky} considered an infinitely thin permanent magnet of finite length. This model is acceptable for calculation of the field at large distances from the source, but it can not be used for the near field calculations.

In paper \cite{Georgiou} has been found an exact special relativistic solution for the electromagnetic field in the interior and  exterior of rapidly rotating perfectly conducting magnetized sphere. The calculation of the field is made as generalization of the field of slowly rotating magnetized neutron star, which was studied in \cite{Rezzolla} under consideration of general relativistic effects. The field of a rotating magnetized sphere which is neither a conductor nor  a dielectric was calculated by Kaburaki \cite{Kaburaki80}.
There is a great variety of other papers which present calculations of the electromagnetic field of  rotating magnetized sphere -- see references in the articles cited above. The results differs essentially dependent on the used model of the magnetized sphere and its speed of rotation.

The model of relativistically rotating sphere is rather complicate. First of all, a solid sphere is  incompatible with the theory  of relativity. Hence, we have to consider a liquid model or gaseous. Therefore it is not a sphere. Secondly, the inner field of fast rotating body depends on the form of the body and on the nature of magnetization. In this paper we accept as a model of the exterior electromagnetic field the field of slowly rotating uniformly magnetized sphere. We show that in the non-relativistic limit the results of many authors cited above give the same field.

Let us start with the expression for the exterior electromagnetic field obtained by Deutsch \cite{Deutsch}. We expand these equations  in powers of  $a=\displaystyle\frac{\omega r_0}{c}$, where $\omega$ is the angular speed of rotation, $r_0$ is the radius of the sphere, and $c$ is the speed of light. But we keep terms like $r_0/r$ which are sufficient near the surface of the sphere.
 Up to the first order of $a$ we receive the next equations for the electric ($\bm E$) and magnetic ($\bm H$) field vectors in a spherical coordinate system $r,\theta,\varphi$ (axis $Z$ is directed along the vector of angular velocity $\bm\omega$):
\begin{eqnarray}
%\label{D1}
E_r&=&-\frac{\mu k^3a^2}{2\rho^4}[\cos\alpha(3\cos2\theta+1)+\sin\alpha\sin2\theta(3C-\rho^2\cos\lambda)], \nn\\
\label{D2}
E_\theta&=&-\frac{\mu k^3}{\rho^2}\left[C\sin\alpha\left(1-\frac{a^2}{\rho^2}\cos2\theta\right)+\frac{a^2}{\rho^2}\cos\alpha\sin2\theta\right], \\
%\label{3}
E_\varphi&=&\frac{\mu k^3}{\rho^2} S\sin\alpha\cos \theta \left(1-\frac{a^2}{\rho^2}\right),\nn
\end{eqnarray}

%Magnetic field vector has components:
\begin{eqnarray}
\label{D4}
H_r&=&\frac{2\mu k^3}{\rho^3} (\cos \alpha\cos \theta%\\
+C\sin \alpha \sin \theta),\nn\\
%\label{D5}
H_\theta&=&\frac{\mu k^3}{\rho^3} [\cos \alpha \sin \theta%\nn\\
 -\sin \alpha\cos \theta  (C-\rho ^2\cos\lambda) ],\\
%\label{6}
H_\varphi &=&\frac{\mu k^3}{\rho^3}\sin \alpha (S-\rho ^2\sin\lambda )\nn\, ,
\end{eqnarray}
where
\[
S=\sin\lambda-\rho\cos\lambda,\quad C=\cos\lambda+\rho\sin\lambda,
\]
$\bm\mu$ is the dipole moment vector, $\mu=|\bm\mu|$, $\lambda=\rho+\varphi-\omega t,\,
\rho=r\omega/c,\, k=\omega/c$, and
 $\alpha$ is the  angle between the vectors $\bm\mu$ and $\bm\omega$.
We have also expanded:
\bea
\sin(\lambda-\beta)\approx\sin\lambda-\beta\cos\lambda,\nn\\
\cos(\lambda-\beta)\approx\cos\lambda+\beta\sin\lambda.\nn
\eea
The magnetic field (\ref{D4}) is the field of rotating point-like magnetic dipole, while the electric field (\ref{D2}) is a superposition of dipole and quadrupole fields. The quadrupole part is presented by terms proportional to $a^2/\rho^2$ and decreases with distance as $\rho^{-4}$.  At great distances $\rho\gg a$ this part vanishes and the electromagnetic field becomes that of rotating magnetic dipole.

The field near the surface of the magnetized body highly depends on the used model. The field (\ref{D2}) is calculated for a  perfectly conducting sphere. The field of an inclined rotator calculated by authors cited above differs substantially from that given by Deutsch \cite{Deutsch}. But if we expand the field obtained for different models  in powers of  $a$,  far from the surface it takes the form of the field of a rotating point like dipole. For example, that is the case for the fields which considered in 
\cite{Ferrari73}, \cite{Kaburaki80}.  Hence, we consider the dipole field as a general case at great distance. Then, the magnetic field is given by Eqs  (\ref{D4}), and the electric field is
\begin{eqnarray}
%\label{D1}
E_r&=&0, \nn\\
\label{D2-1}
E_\theta&=&-\frac{\mu k^3}{\rho^2}C\sin\alpha, \\
%\label{3}
E_\varphi&=&\frac{\mu k^3}{\rho^2}S \sin\alpha\cos \theta.\nn
\end{eqnarray}
The fields (\ref{D4}) and  (\ref{D2-1}) can be represented by 4-dimensional vector potential $A^\nu$. In the spherical coordinate system $x^\nu=(ct,r,\theta,\varphi)$ it is
\bea\label{potent}
&&A^0=A^1=0,\nn\\
&&A^2=-\frac{\mu}{r^3}S\sin \alpha,\\
&&A^3=\frac{\mu}{r^3}(\cos\alpha-C\sin \alpha\cot\theta).\nn
\eea

In the next three sections we study dynamics and the potential energy for the charged particles in the field of rotating dipole given by Eqs   (\ref{D4}) and  (\ref{D2-1}), and in Section \ref{sec-surf} we describe the potential energy near the sphere surface, according to Eqs  (\ref{D2}) and  (\ref{D4}).
%%%%%%%%%%%%%%%%%%%%%%%%%%
%%
\section{Integral of motion for the particles in arbitrary rotating electromagnetic  field}\label{sec3}
%%
%%%%%%%%%%%%%%%%%%%%%%%%%%
Let us consider an arbitrary electromagnetic field rotating with angular velocity $\omega$.
The four-dimensional potential of such field in the inertial spherical coordinate system $x^\nu=(ct,r,\theta,\varphi),\, \nu=0,1,2,3$ is defined as
\[A^\nu=A^\nu(r,\theta,\varphi-\omega t+\rho).\]

In the corotating reference frame $x^{\nu'}=(ct,r,\theta,\psi)$, with $\psi=\varphi-\omega t$, the
field does not depend on time.
Hence, the corresponding generalized momentum is conserved.
The Lagrangian for a charged particle with mass $m$ and charge $e$ in rotating reference frame is
\be\label{Lag}
L=\frac{m}{2}u^{\nu'}u_{\nu'}+\frac{e}{c} u_{\nu'}A^{\nu'},\quad u^{\nu'}=(c\dot{t}, \dot{r},\dot{\theta},\dot{\psi}),
\ee
where $u^\nu$ is the four-dimensional velocity, prime shows that the quantity relates to the rotating reference frame, and the dot denotes derivative with respect to the proper time $\tau$.
As stated above, the time component  $p_ {0'}$ of the generalized 4-momentum is an integral of motion:
\be
\label{pnul}
p_{0'}=\frac{\partial L}{\partial u^{0'}}= m u_{0'}+\frac{e}{c}A_{0'}.
\ee
This means that the energy of the particle in the corotating frame defined as ${\cal E}'=cp_{0'}$ is conserved.

Let us express the integral of motion $P=p_{0'}$ in terms of quantities in the inertial reference frame. The matrix of transformation from the inertial reference frame to the rotating one reads
\be\label{matJ}
J_{\mu'}^{\phantom{\mu}\nu}=\frac{\partial x^{\nu}}{\partial x^{\mu'}}=
\begin{pmatrix}
1&0&0&\omega/c\\
0&1&0&0\\
0&0&1&0\\
0&0&0&1
 \end{pmatrix}
\ee
As a result of transformations
$
A_{\mu'}=J_{\mu'}^{\phantom{\mu}\nu}A_\nu
$
we obtain
\be\label{intR}
P=p_0+\frac{\omega}{c}p_3,
\ee
where $p_\nu=mu_\nu+\frac ec A_\nu$ are the generalized momenta in the inertial frame. The quantity $p_3$ is the generalized angular momentum relative to the axis of rotation. If we multiply Eq. (\ref{intR}) by $c$, we can read it as:
 the sum of energy and angular momentum, multiplied by $\omega$, is conserved.

Let us calculate the integral of motion for the field of  precessing dipole. The generalized momenta calculated by use of Eqs  (\ref{potent}) are:
\bea
p^0&=&mc\dot t,\\
p^3&=&m\dot\varphi+\frac{e\mu}{cr^3}(\cos\alpha-C\sin \alpha\cot\theta).
\eea
Using the metric tensor for the spherical coordinates in the inertial frame of reference
\[
g_{\mu\nu}={\rm diag} (1,-1,-r^2, -r^2\sin^2\theta)
\]
we find the integral of motion
\bea\label{p-inert}
P=m(c\dot t -\frac\omega c r^2\dot\varphi\sin^2\theta)-\frac{e\mu\omega}{c^2r}\sin\theta(\cos\alpha\sin\theta-C\sin\alpha\cos\theta).
\eea
Substituting  $\varphi=\psi+\omega t$ we obtain the expression for $P$ in the corotating reference system:
\bea\label{int1}
P&=&m[c\dot t(1-\rho^2\sin^2\theta) -\frac{\omega r^2}{c}\dot\psi\sin^2\theta]-\frac{e\mu\omega}{c^2r}\sin\theta\{\cos\alpha\sin\theta-[\cos(\rho+\psi)+\\
&+&\rho\sin(\rho+\psi)]\sin\alpha\cos\theta\}.
\eea
If we consider  $r,\theta,\psi$ as the particle coordinates, then the expression (\ref{int1}) is valid only inside the light cylinder of radius $c/\omega$, while Eq. (\ref{p-inert}) is  correct throughout the entire space.

%%%%%%%%%%%%%%%%%%%%%%%%%%%%
%%
\section{Potential energy}\label{sec4}
%%
%%%%%%%%%%%%%%%%%%%%%%%%%%%%%

We study the particle dynamics with respect to the rotating reference frame with coordinates $ct,r,\theta,\psi$ and the metric defined by tensor
\be\label{metric-pr}
g_{\mu'\nu'}=
\begin{pmatrix}
1-\rho^2\sin^2\theta&0&0&- r\rho\sin^2\theta\\
0&-1&0&0\\
0&0&-r^2&0\\
-r\rho\sin^2\theta&0&0&-r^2\sin^2\theta
 \end{pmatrix}
\ee

The total energy $\cal E'$ of a particle in curved space can be expressed as follows \cite{Landau}, 
%\S 88
\be
{\cal E'}=cp_{0'}=\frac{mc^2\sqrt{g_{0'0'}}}{\sqrt{1-\beta^2}}+eA_{0'},
\ee
where $\beta=v/c$, and $v$ is the particle velocity. As $mc^2\sqrt{g_{0'0'}}$ is the energy of the particle at rest, we can define the kinetic energy as
\be
T=mc^2\sqrt{g_{0'0'}}\left(\frac{1}{\sqrt{1-\beta^2}}-1\right)
\ee
Then, the  potential energy $U$ can be introduced as $U=cp_{0'}-T$, which gives
\be\label{Vef}
U=mc^2\sqrt{g_{0'0'}}+eA_{0'}
\ee
The potential energy defined by Eq. (\ref{Vef}) possesses a standard property: the space part of its four dimensional gradient $\partial_\nu U$ is proportional to acceleration of the particle being at rest. In order to prove this we consider equation of motion
\be\label{mot1}
m\dot u_\nu+\frac ec u^\sigma F_{\sigma\nu}-\frac m2 u^\sigma u^\rho \partial_\nu g_{\sigma\rho}=0,
\ee
where $F_{\sigma\nu}=\partial_\sigma A_\nu-\partial_\nu A_\sigma$. Substituting the four-velocity $u^{\sigma'}=(u^{0'},0,0,0)$ for the particle at rest we obtain
\[
m\dot u_{\nu'}=u^{0'}\left(\frac ec \partial_{\nu'}A_{0'}+\frac m2 u^{0'}\partial_{\nu'}g_{0'0'}\right),
\]
It follows from $u^{\sigma'}u_{\sigma'}=c^2$ that $u^{0'}=c/\sqrt{g_{0'0'}}$. Hence,
\be\label{grad}
m\dot u_{\nu'}=\frac{1}{\sqrt{g_{0'0'}}} \partial_{\nu'}\left(eA_{0'}+ mc^2 \sqrt{g_{0'0'}}\right),
\ee
which proves the statement.

Let us find the potential energy of a particle in the rotating dipole field. Transformation of the potential (\ref{potent}) into rotating reference frame by use of matrix (\ref{matJ}) leads to
\[
A_{0'}=\frac{\mu\omega}{2rc}[\sin\alpha\sin 2\theta(\cos\xi+\rho\sin\xi)-2\cos\alpha\sin^2\theta],
\]
where $\xi=\rho+\psi$.
Substituting this into Eq. (\ref{Vef}) and introducing a dimensionless potential energy $V=U/mc^2$ we obtain
\bea\label{Vpotent}
V=\sqrt{1-\rho^2\sin^2\theta}+\frac{N_\perp}{2\rho}\sin 2\theta(\cos\xi+\rho\sin\xi)-
\frac{N_\parallel}{\rho}
\sin^2\theta
\eea
with
\be\label{defN}
N_{\perp}=N\sin\alpha,\quad N_{\parallel}=N\cos\alpha,
\quad N=\displaystyle\frac{e\mu\omega^2}{m c^4}.
\ee
Notice, that all physical parameters are gathered in one dimensionless parameter $N$. For example, for electrons the value of $N$ for pulsar in Crab Nebula is $5\cdot10^{10}$, Jupiter -- $0.03$, Earth -- $3\cdot10^{-7}$ and for the magnetized sphere used in experiment by \cite{Timofeev} -- $3\cdot10^{-16}$.

As shown in Paper I, in case of $N\ll 1$ the potential energy changes sufficiently in the region $\rho\sim N^{1/3}\ll 1$.
Expanding (\ref{Vpotent}) in power series in $\rho$ we receive the  effective potential energy studied in details  in Paper I. It was shown there that the value of $N$ plays the role of a scale factor and in terms of reduced variable $\tilde\rho=\rho N^{-1/3}$ the shape of  potential energy does not depend on $N$. In  general case, the domain of potential energy is explicitly restricted by the light cylinder $\rho^2\sin^2\theta<1$, as  indicated by Eq. (\ref{Vpotent}).
%%%%%%%%%%%%%%%%%%%%%%%%%%%%%%%%%%%%%%%%%%
%%
\section{Equipotential surfaces}\label{sec5}
%%
%%%%%%%%%%%%%%%%%%%%%
In this section we present the profiles of potential energy defined by Eq. (\ref{Vpotent}). Due to the argument $\xi=\rho+\psi$ the equipotential surfaces take a form of surfaces twisted around the $Z$-axis. If we ``twist back'' the whole picture, introducing coordinate $\eta=\psi+\rho-\sigma$, with
\be
\sin\sigma=\frac{\rho}{\sqrt{1+\rho^2}},\quad \cos\sigma=\frac{1}{\sqrt{1+\rho^2}}.
\ee
we find out that the potential energy becomes symmetric with respect to the plane $\eta=0,\,\pi$ which contains vectors $\bm\mu$ and $\bm\omega$:
\bea\label{Vpotent-eta}
V=\sqrt{1-\rho^2\sin^2\theta}+\frac{N_\perp}{2\rho}\sqrt{1+\rho^2}\sin 2\theta\cos\eta-
\frac{N_\parallel}{\rho}
\sin^2\theta.
\eea
Besides, the function $V$ is symmetric with respect to transformations $\eta\to\eta+\pi; \theta\to\pi -\theta$.

At the plots below we show the  sections of the equipotential energy surfaces  by plane $\eta=\rm const$. The equipotential surfaces are marked by  numbers equal to the energy level $V=\rm const$. It means that the particle having the total energy $\cal E'$  has zero velocity at  equipotential surface $V={\cal E'}/mc^2$ and can move according to the equations of motion  in the area where the potential energy is less the its total energy. For example, a particle with total energy ${\cal E'}=1.04mc^2$  being in the field depicted in Fig. \ref{fig1} can move everywhere except the closed region at the centre marked by number 1.04.

We study the structure of potential energy for $\alpha\le\pi/2$, for positive and negative charge of the particles. The structure for  $\alpha>\pi/2$ is the same but $e$ should be replaced by $-e$ and $\theta$ by $\pi-\theta$.
All profiles  are plotted for the inclination angle $\alpha=60^0$, if other angle $\alpha$ is not specified explicitly.
%%%%%%%%%%%%%%%%%%%%%%
%%
\subsection{Equipotential surfaces for positively charged particles}
\begin{figure}[htbp]
\begin{center}
\includegraphics[width=2.3in]{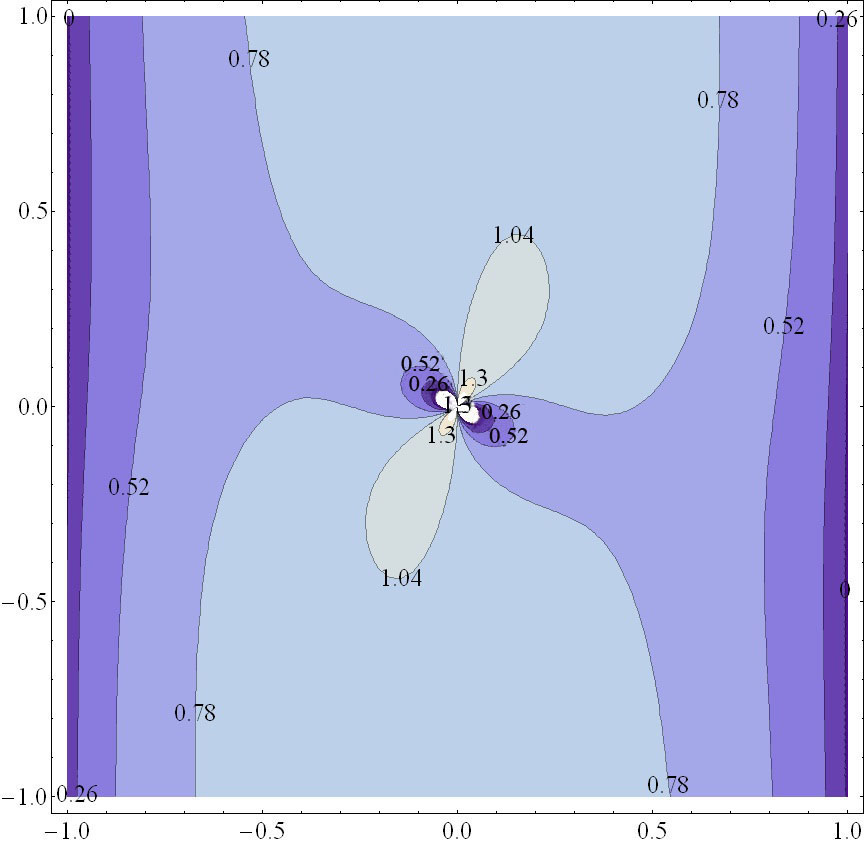}
\caption{Equipotential profiles for $ N=0.1, \eta=0$.}
\label{fig1}
\end{center}
\end{figure}

\begin{figure}[htbp]
\begin{center}
\includegraphics [width=2.3in]{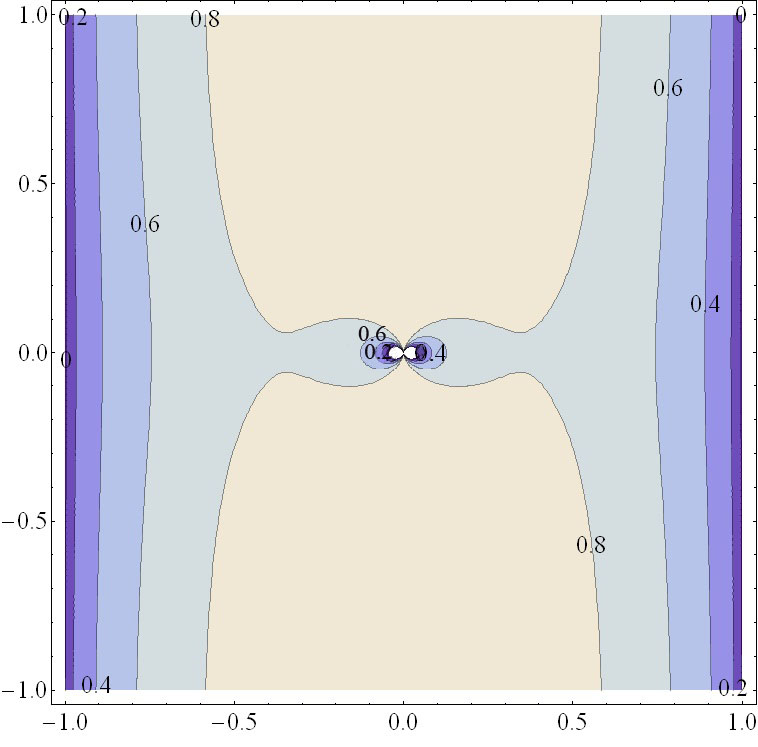}
\caption{Equipotential profiles for $ N=0.1, \eta=90^{0}$}
\label{fig2}
\end{center}
\end{figure}

\begin{figure}[htbp]
\begin{center}
\includegraphics [scale=0.35]{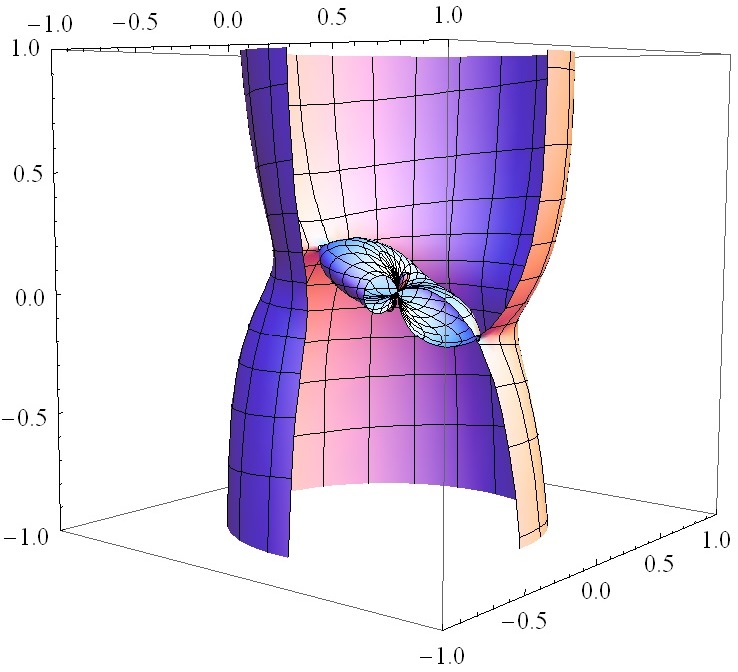}
\caption{Equipotential surface   for $ N=0.1$, $V=0.74$}
\label{fig3}
\end{center}
\end{figure}

%%
%%%%%%%%%%%%%%%%%%%%%%
We start with equipotential surfaces for small $N$.  Fig.~\ref{fig1} and Fig.~\ref{fig2} show the profiles for $N=0.1$. The sign of $N$ is the sign of the particle charge according to definition of $N$ (\ref{defN}).
The equipotential surfaces for small $N$ are almost the same as plotted in Paper I in 3-D form.
For example, the equipotential surface for  $V= 0.74$ is shown in Fig. \ref{fig3}. The constant $C$ used in Paper I and the energy level $V$ of this paper are bound by relation $V=1+N^{2/3}C/2$. The shape of equipotential surfaces at $N\ll 1$ does not depend on $N$. The value of $N$ plays a role of scale factor in  form $V\sim N^{2/3}$.

\begin{figure}[htbp]
\begin{center}
\includegraphics [width=2.3in]{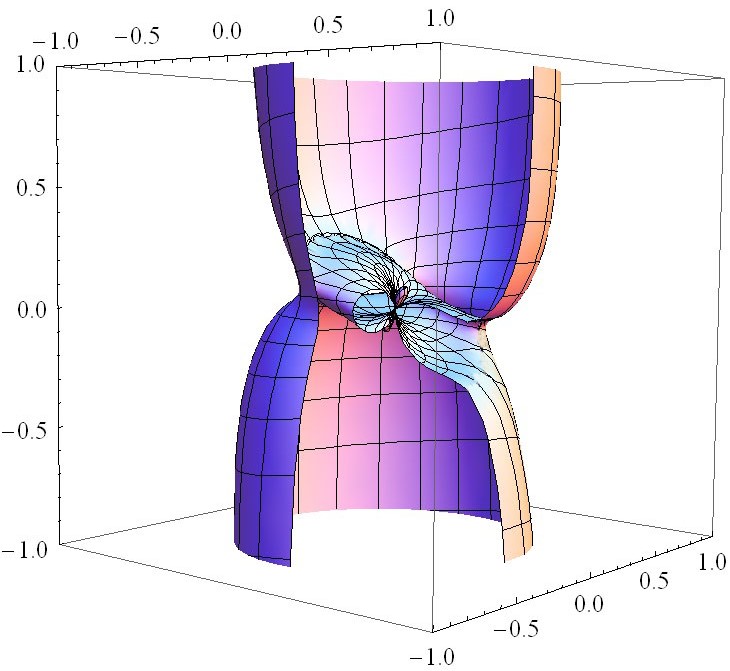}
\caption{Equipotential surface for $ N=0.1, V=0,85$}
\label{fig4}
\end{center}
\end{figure}

 One can see in  Figs. \ref{fig1}--\ref{fig3} that the energy levels form a potential valley in a shape of torus around the centre of the field. 
There are two allowed regions for the particles of energy less than $\approx 0.74$: one is the closed region inside the torus and another is the region outside the cylinder-like surface. At the critical energy level $V\approx 0.74$ the inner and outer regions touch one another at two symmetric  points which are the saddle points  of the potential energy.  This is demonstrated by  Fig. \ref{fig3}.
 It was proved in Paper I that in approximation of small $N$ all the stationary points defined by equations $\partial V/\partial q_i=0$ with $q_i=\rho,\theta,\psi$ are the saddle points. There are also two saddle points in equatorial plane with coordinates defined by Eq. (\ref{7}).  For the particles of energy greater than the critical energy, the inner and outer allowed regions are united by two symmetric conjugation tubes as one can see in Fig. \ref{fig4}. Such particles can escape from the torus-like trapping region to outer space.

\begin{figure}[htbp]
\begin{center}
\includegraphics [width=2.3in]{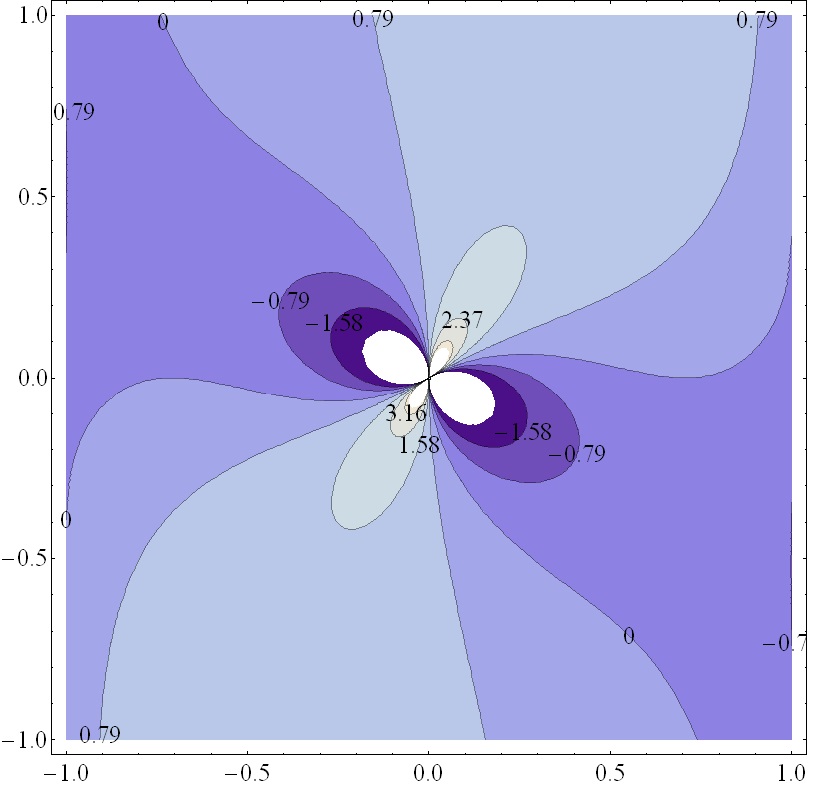}
\caption{ Equipotential profiles for $ N=1, \eta=0$}
\label{fig5}
\end{center}
\end{figure}

\begin{figure}[htbp]
\begin{center}
\includegraphics[width=2.3in]{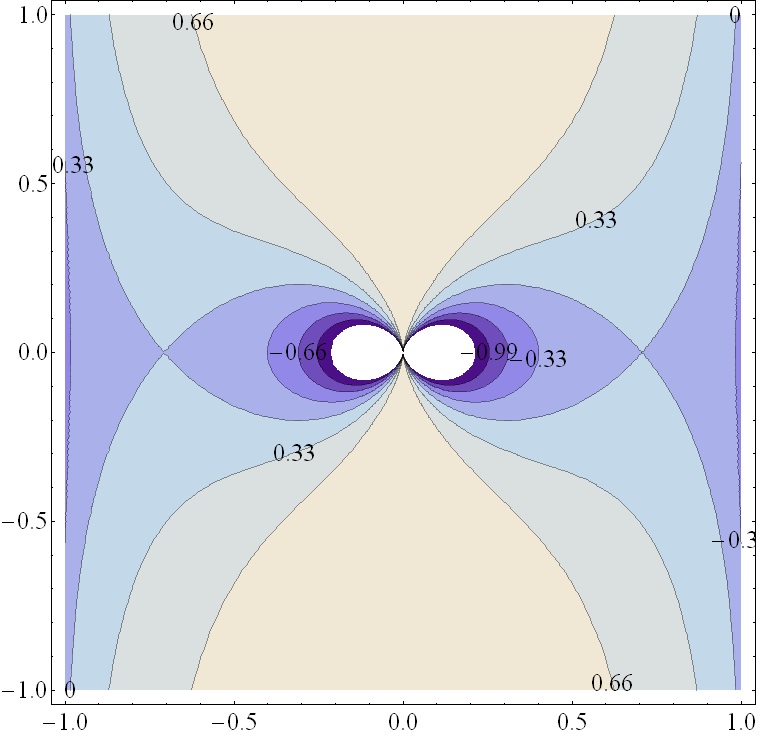}
\caption{Equipotential profiles for $ N=1, \eta=90^{0}$}
\label{fig6}
\end{center}
\end{figure}

\begin{figure}[htbp]
\begin{center}
\includegraphics [scale=0.28]{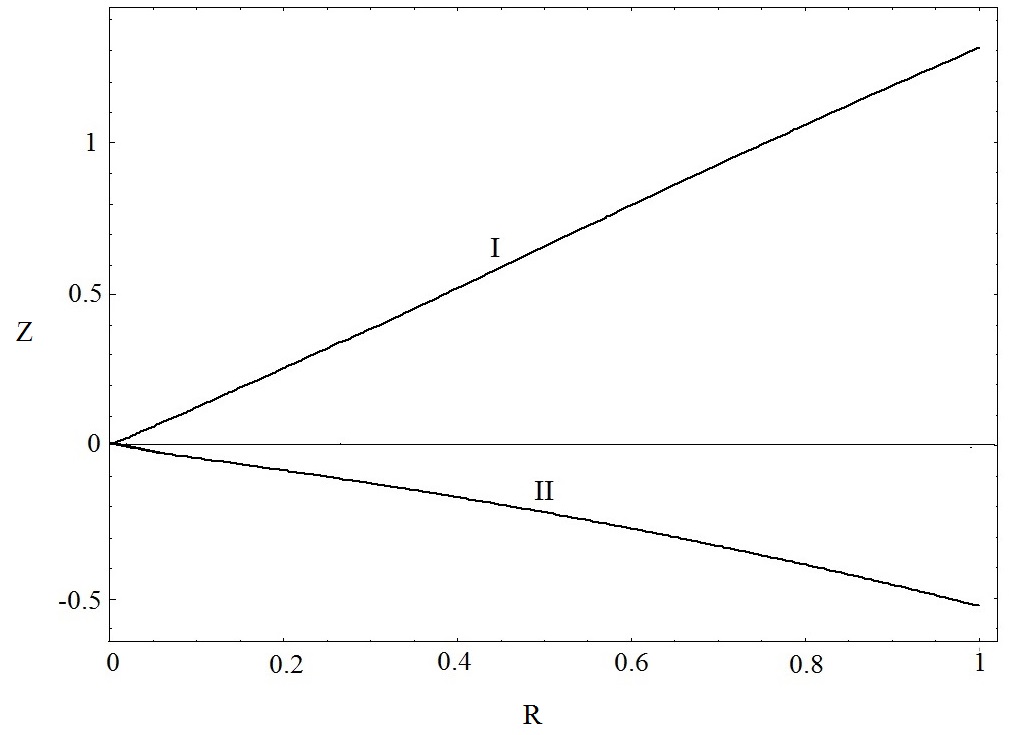}
\caption{Lines along which the saddle points moves as $N$ varies. I -- for negatively and II -- for positively charged particles. $\eta=0,\,\alpha=\pi/3$.}
\label{fig7}
\end{center}
\end{figure}

The potential profiles for intermediate values of $N$ have similar structure. For example, the sections of equipotential surfaces for $N=1$ are depicted in  Figs \ref{fig5}--\ref{fig6}.
The saddle points for this $N$ lie at the energy level $V\approx -0.35$. As $N$ increases, the saddle points move to the light cylinder along a lines given by Eq. (\ref{line}). The line lying in the plane $\eta =0$ is shown in Fig. \ref{fig7} as line II in the 
cylindrical coordinates $R=\rho\sin\theta$ and $Z=\rho\cos\theta$. The corresponding lines in the plane $\eta=\pi$ can be produced by substitution $Z\to -Z$.  It follows from Eqs ({\ref{Ap1}--\ref{Ap2}) that the saddle points coordinates $\rho\sin\theta\to 1$  as $N\to\infty$. I.e. the  saddle points approach the light cylinder asymptotically when $N$ grows. In other words, the critical potential surface, and hence, the trapping regions  exist at any large $N$.
\begin{figure}[htbp]
\begin{center}
\includegraphics [width=2.3in]{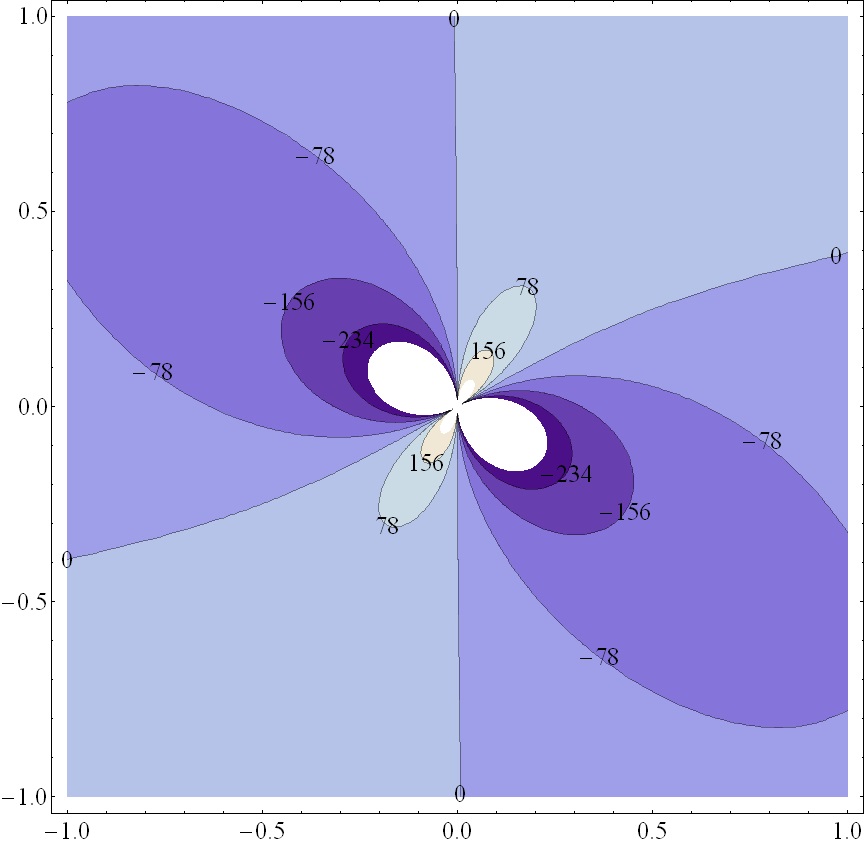}
\caption{Equipotential profiles for $ N=100, \eta=0$}
\label{fig8}
\end{center}
\end{figure}
\begin{figure}[htbp]
\begin{center}
\includegraphics [width=2.3in]{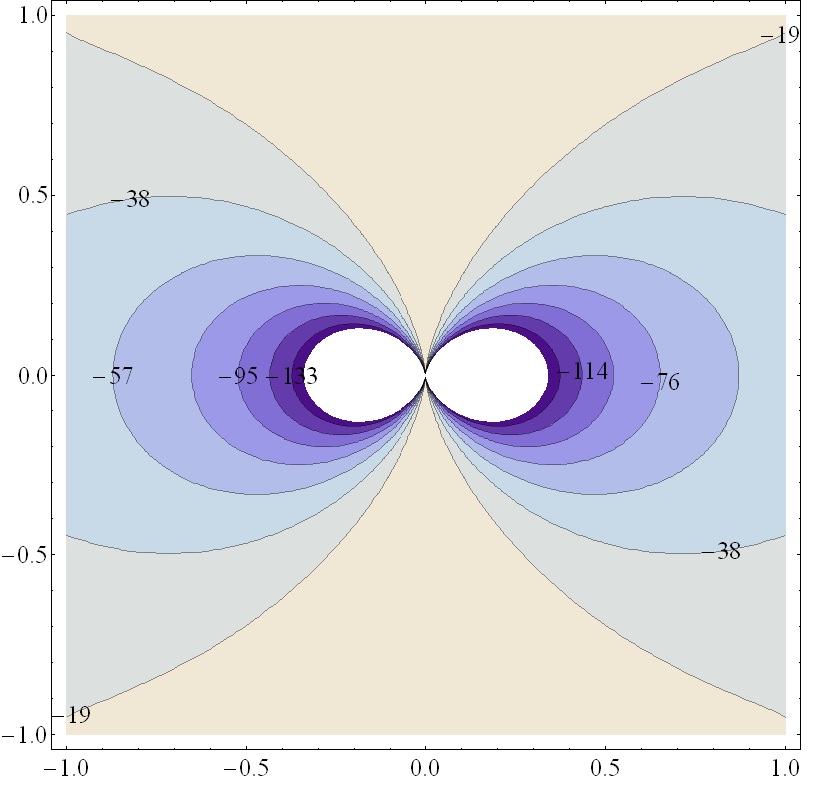}
\caption{Equipotential profiles for $ N=100, \eta=90^{0}$}
\label{fig9}
\end{center}
\end{figure}

Sections of the potential surfaces for $N=100$ are plotted in Fig \ref{fig8} and Fig \ref{fig9}. The shape of the profiles in case of large $N$ does not depend on $N$. Indeed, if $N\gg 1$, we can neglect the first term in Eq. (\ref{Vpotent-eta}) and $N$ becomes just a scale factor. The sole exception is the vicinity of the light cylinder, because the first derivatives $\partial V/\partial \rho$ and $\partial V/\partial\theta$, as one can see in Eqs (\ref{Ap1}) and (\ref{Ap2}), tend to infinity as $\rho\sin\theta\to 1$.
%%%%%%%%%%%%%%%%%%%%%%%%%%%%%%%
%%
\subsection{Equipotential surfaces for negatively charged particles}
%%
%%%%%%%%%%%%%%%%%%%%%%%%%%%%%%%
There is a significant difference between  the structure of equipotential surfaces for positive and negative charges, though they share a number of traits. As we change the sign of the charge in expression for potential energy, the ``potential hills'' become ``potential valley'' and vice versa.  The  trapping regions in this case have a form of two symmetric dumb-bell shaped figures. 
The sections of level surfaces for $N=-1$ are shown in Figs \ref{fig10}  and \ref{fig11}. 
\begin{figure}[htbp]
\begin{center}
\includegraphics [width=2.3in]{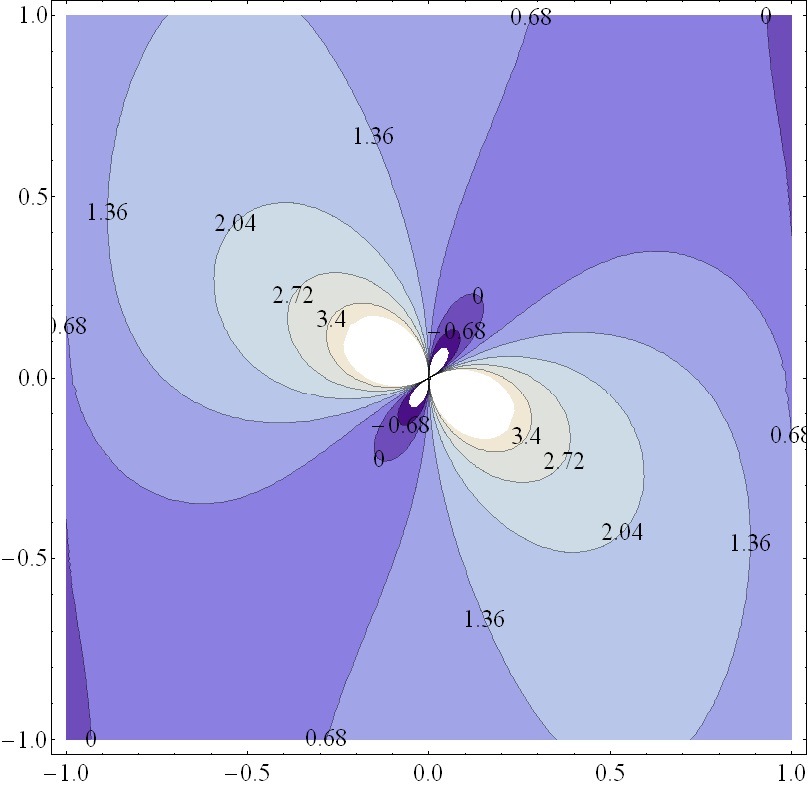}
\caption{Equipotential profiles for $ N=-1, \eta=0$.}
\label{fig10}
\end{center}
\end{figure}
\begin{figure}[htb]
\begin{center}
\includegraphics [width=2.3in]{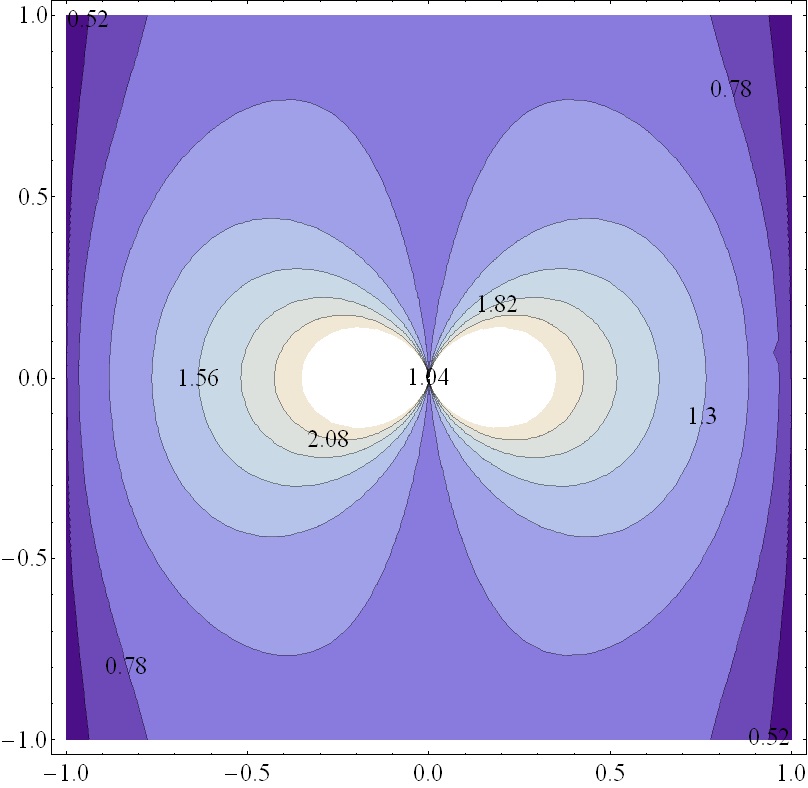}
\caption{Equipotential profiles for $ N=-1, \eta=90^0$.}
\label{fig11}
\end{center}
\end{figure}
\begin{figure}[htb]
\begin{center}
\includegraphics [scale=0.4]{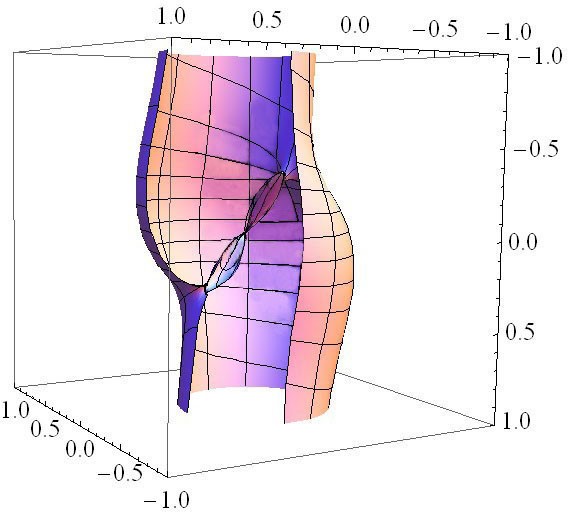}
\caption{Equipotential surface   for $ N=- 0.1$, $V\approx0.82$.}
\label{fig12}
\end{center}
\end{figure}
 For the negatively charged particle  there is still a critical energy level at which the inner trapping region contacts the outer cylinder-like surface at two saddle points. The shape of the critical surface which is of level  $V\approx0.82$ is depicted in Fig. \ref{fig12}. 

As $N$ varies, the saddle point in the plane $\eta=0$ moves along the line I of Fig. \ref{fig7} and approaches the light cylinder of radius $R=\rho\sin\theta=1$ as $N\to\infty$. The lines I and II in Fig. \ref{fig7} intersect at the coordinate origin at angle $\gamma$: $\cos\gamma=\frac 13 \sin\alpha$.

%%%%%%%%%%%%%%%%%%%%%%%%%%%%
%%
\subsection{Orthogonal rotator}
%%
%%%%%%%%%%%%%%%%%%%%%%%%%%%%%%%

In this section we describe shortly the structure of potential energy for the  inclination angle $\alpha=\pi/2$. In this case the last term in Eq. (\ref{Vpotent-eta}) vanishes. Hence the equipotential surfaces for positive and negative charges become symmetric because substitution $e\to-e$ is equal to substitution $\eta\to\eta+\pi$. 

The energy profiles for negative charge do not change sufficiently as $\alpha$ tends to $\pi/2$. But the profiles for positive charge change  substantially as one can see in Figs  \ref{fig13} and  \ref{fig14}.

The specific profiles of Fig.  \ref{fig14} arise as a consequence that the saddle points in equatorial plane,  seen for example in Fig. \ref{fig6},  move to $Z$ axis according to Eq. (\ref{7}), as the angle $\alpha$ approaches to $\pi/2$. And all the central pattern shrinks to the coordinate origin.
The profiles for negative charge are the same, but rotated around $Z$-axis through $180^0$. The trapping regions both for negative and positive charges  form a dumb-bell shaped figures.
\begin{figure}[htb]
\begin{center}
\includegraphics [width=2.3in]{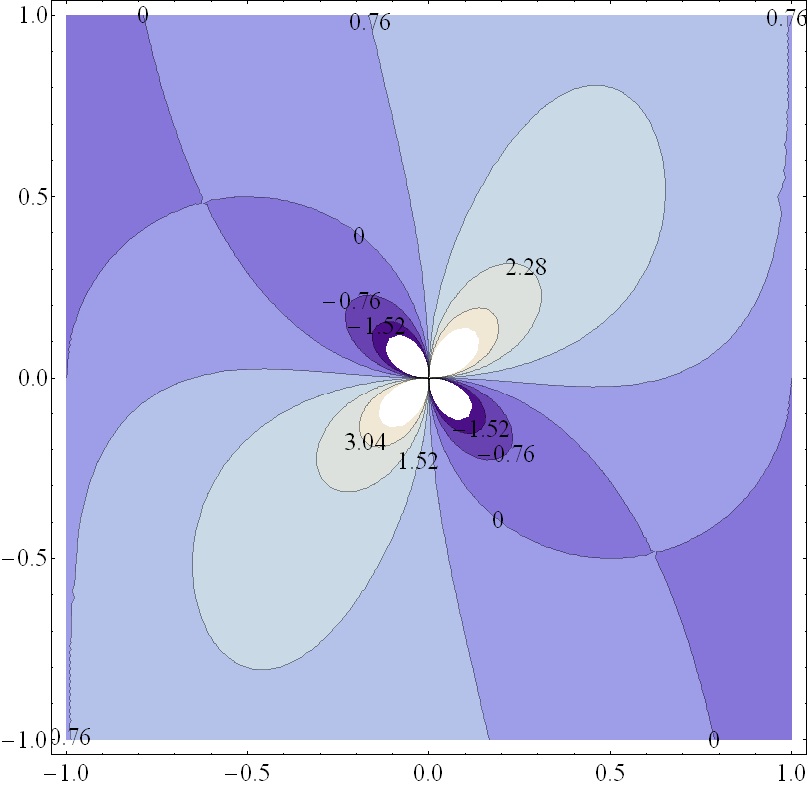}
\caption{Equipotential profiles for $ N=1,  \eta=0, \alpha=\pi/2$.}
\label{fig13}
\end{center}
\end{figure}
\begin{figure}[htb]
\begin{center}
\includegraphics [width=2.3in]{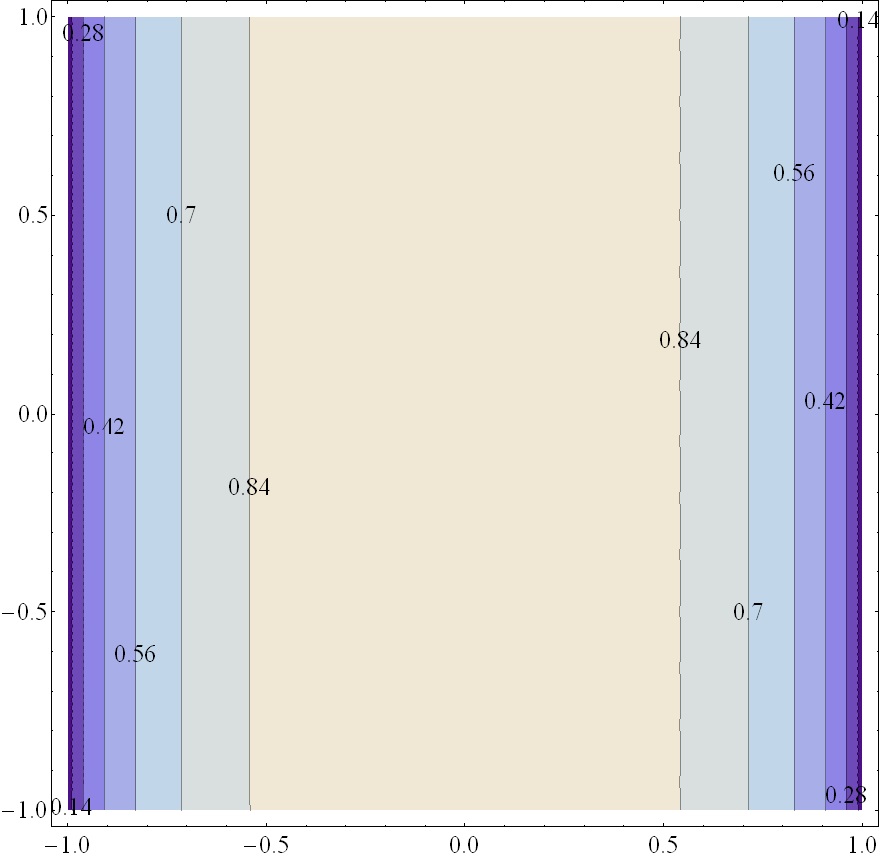}
\caption{Equipotential profiles for $ N=1,  \eta=90^0, \alpha=\pi/2$.}
\label{fig14}
\end{center}
\end{figure}

%%%%%%%%%%%%%%%%%%%%%%%%%%%%%%%%%%
%%
\section{Potential energy near the surface of uniformly magnetized sphere}\label{sec-surf}
%%
%%%%%%%%%%%%%%%%%%%%%%%%%%%%%%
Up to now we have studied the particles dynamics on a large scale, up to the light cylinder. The constructed plots are not valid near the surface of a star or a planet because the quadrupole electric field is neglected. As discussed in Sec. \ref{sec2}, the electric field not far from the surface essentially depends on the used model. In this section we investigate the potential energy of charged particles in the field of perfectly conducting sphere found in \cite{Deutsch}. These fields are described by Eqs  (\ref{D2}) and (\ref{D4}). The respective 4-dimensional vector potential  can be written as follows
\bea\label{pot-sur}
&&A^0=-\frac{\omega r_0^2\mu}{6 c r^3}\left(3C\sin2\theta\sin\alpha+\cos\alpha(3\cos2\theta+1)\right),\nn\\
&&A^1=0,\nn\\
&&A^2=-\frac{\mu}{r^3}S\sin \alpha,\\
&&A^3=\frac{\mu}{r^3\sin\theta}(\cos\alpha\sin\theta-C\sin \alpha\cos\theta),\nn
\eea
Transforming this potential to rotating reference frame and substituting it  into Eq. (\ref{Vef}) we obtain the potential energy with regard to the quadrupole electric field
\begin{multline}\label{Vpotent-1}
V=\sqrt{1-\rho^2\sin^2\theta}+\left[\frac{N_\perp}{2\rho}\sin 2\theta(\cos\xi+\rho\sin\xi)-\frac{N_\parallel}{\rho}
\sin^2\theta\right]\left(1-\frac{a^2}{\rho^2}\right)-\frac{2N_{\parallel}}{3\rho}\frac{a^2}{\rho^2}.
\end{multline}
It differs from the potential energy (\ref{Vpotent}) by terms proportional to $a^2/\rho^2$.

The magnitude of $a$ for real celestial bodies is well below unity. For example, the values of $a$  for Earth, Jupiter and pulsar in Crab Nebula are $1.5\cdot10^{-6}$, $4\cdot10^{-5}$ and $7.6\cdot10^{-3}$ respectively. We have plotted the equipotential surfaces for $a=0,01$.  Built for the interval $0<\rho\sin\theta<1$, the graphs do not differ essentially from those presented in the previous section, except for details at the coordinate origin. These details are shown in Figs \ref{fig15} -- \ref{fig18}.
\begin{figure}[htbp]
\begin{center}
\includegraphics [width=2.3in]{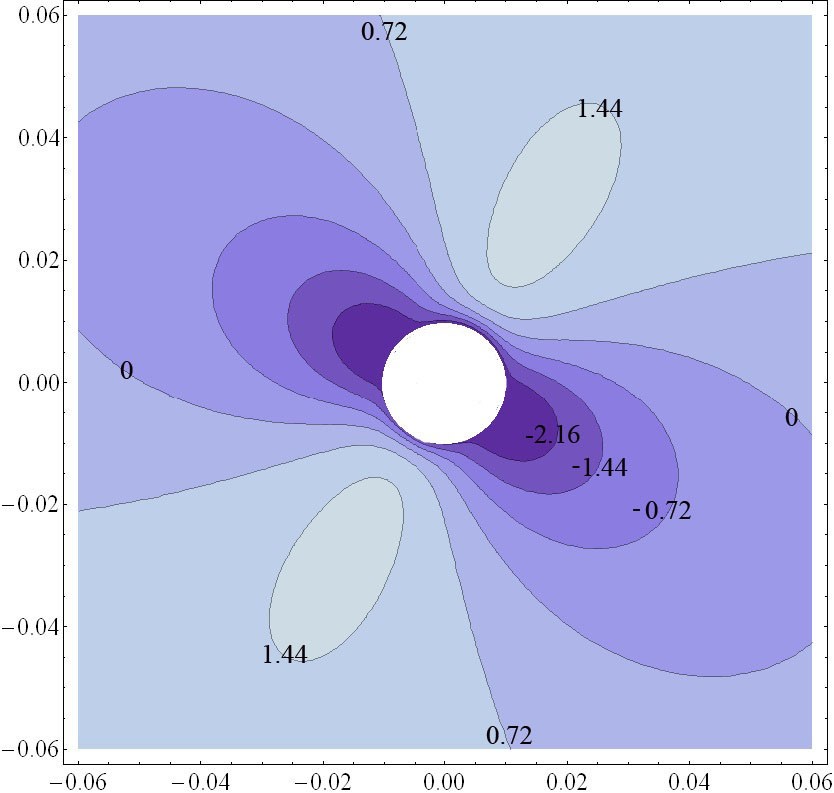}
\caption{Equipotential profiles for $ N=0.1,  \eta=0$.}
\label{fig15}
\end{center}
\end{figure}
\begin{figure}[htbp]
\begin{center}
\includegraphics [width=2.3in]{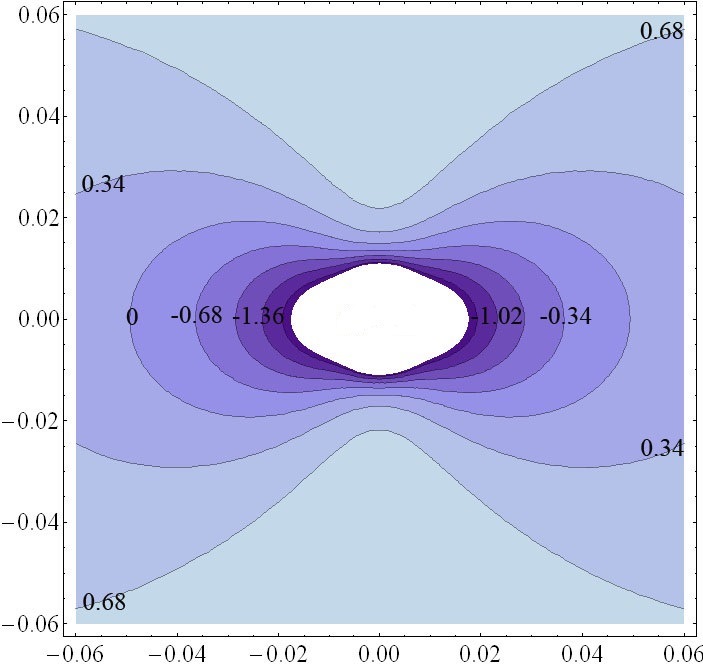}
\caption{Equipotential profiles for $ N=0.1,  \eta=90^0$.}
\label{fig16}
\end{center}
\end{figure}
\begin{figure}[htb]
\begin{center}
\includegraphics [width=2.3in]{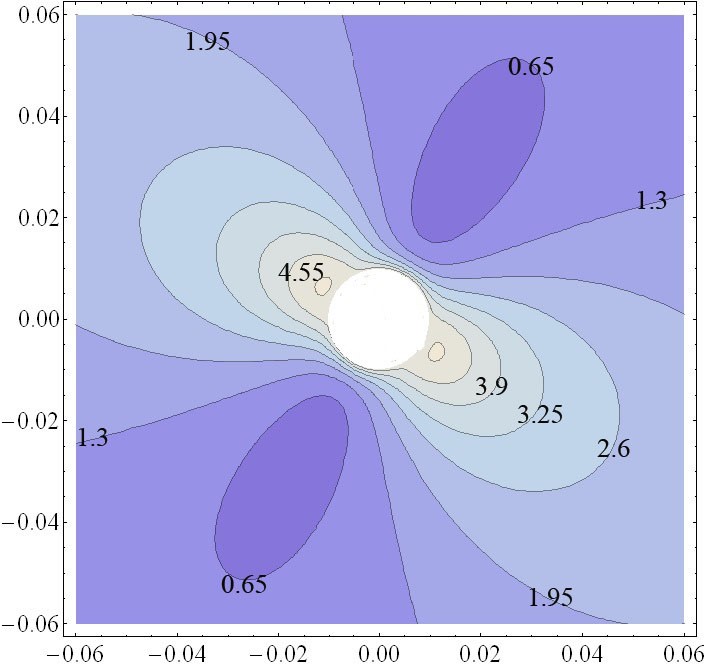}
\caption{Equipotential profiles for $N=- 0.1,  \eta=0$.}
\label{fig17}
\end{center}
\end{figure}
\begin{figure}[htb]
\begin{center}
\includegraphics [width=2.3in]{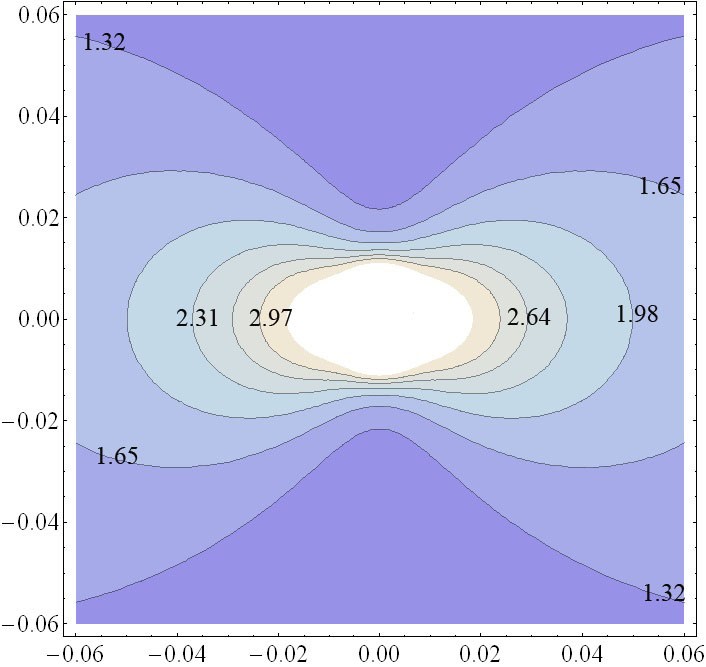}
\caption{Equipotential profiles for $N=- 0.1,  \eta=90^0$.}
\label{fig18}
\end{center}
\end{figure}

The distinctive property of potential energy in this case is that there is a minimum of the potential energy for negatively charged particles, and the trapping region is separated from the surface of the sphere as shown in Fig. \ref{fig17}. The trapping region for the positively charged particles still forms of a torus encircling the star and siding the star surface along the magnetic equator (Figs \ref{fig15} and \ref{fig16}).
The shape of the profiles within the area $\rho\sim a$ does not change sufficiently as $N$ varies. The reason is that the first term in Eq. (\ref{Vpotent-1}) is close to unity in case of small $\rho$, and it can be neglected as a constant in the potential energy. Hence, $N$ becomes just a scale parameter.

%%%%%%%%%%%%%%%%%%%%%%%%%%%%%
%%
\section{Discussion}\label{sec6}
%%
%%%%%%%%%%%%%%%%%%%%%%%%%%%%%%

The potential energy of Eq. (\ref{Vpotent-eta}) was constructed under supposition that the magnetized body is in vacuum and there are not regions with non zero net charge in the surrounding plasma. But we see that the trapping regions for particles of different charge are located in different areas. Hence, this can cause  sufficient charge separation in the magnetosphere. If say, a pair of particles is born at some point as a result of pair production,  one of the particles find itself in the potential well while the other one on the potential hill. As a consequence the first one can be trapped in the region while the particle with opposite charge will leave the region with acceleration. As the trapping regions accumulate sufficiently great charge, the potential profiles will be distorted. In this case we have to study the plasma dynamics rather than motion of a single particle. A great variety of papers on this subject are cited in \cite{Michel} and in recent review \cite{Beskin}.

In the field with large $N$ such as of neutron star, the relativistic charged particles undergo intensive radiative friction force,
 leading to particle energy losses. Though, the particle motion with respect to the radiation reaction is not a subject of this paper, the potential energy study is nevertheless, a powerful instrument of qualitative analysis of the particle behaviour in this case. Because the structure of the potential energy is defined solely by the field and does not depend on the particle motion. It is obvious that if the particle loses its energy due to radiation, it  progressively passes  to lower energy levels continuing its motion along other magnetic field lines. Hence, the boundary restricting the allowed region changes with time, shifting the particle downhill the potential profile.     If the particle in this motion encounters the force-free surface defined by $\bm E\cdot \bm H=0$, it can  join the surface after some oscillations, and then drift along the surface as long as its energy is conserved \cite{Jackson, Istomin}, but still within the area bounded by appropriate equipotential surface.  
As may be seen from the figures above, the trapped positively charged particles eventually  fall on the star surface as they lose their energy, while the others can move off to infinity.  The particle orbits with  regard to radiation reaction have been numerically calculated in the field of orthogonal rotating dipole \cite{Laue}. It was shown that there is a critical surface such that the trajectories starting inside the surface end on polar regions, and the outside trajectories recede to infinity. 

The fact that in case $\bm \omega\cdot\bm \mu>0$ the negatively charged particles concentrate in polar regions of inclined rotator and the particles with positive charge in equatorial zone, coincides with conclusions made by other authors, who have used different models for the neutron star magnetosphere \cite{Gold, Jackson, Istomin}.

\section*{Acknowledgement}
This research has been supported by the grant for LRSS, project No 88.2014.2

\appendix
\section{Stationary points of the relativistic potential energy}
%%
%%%%%%%%%%%%%%%%%%%%%%%%%
The power of potential formulation of the problem is the possibility to find the ``potential valleys'' where the charged  particles can be trapped. And the slope of the ``valley'' shows the force exerted on the particle.  Having this in mind, we find the stationary points of the potential energy, i.e. the points satisfying  the set of equations:
\[\frac{\partial V}{\partial q_i}=0,\]
where $q_i=\rho,\theta,\psi$.
This gives a system of three equations
\bea
\label{Ap1}
&&\frac{\rho^3\sin\theta}{\sqrt{1-\rho^2\sin^2\theta}}+\frac{N_\perp\cos\theta}{\sqrt{1+\rho^2}}(\cos\eta+\rho^3\sin\eta)-N_\parallel\sin\theta=0,\\
\label{Ap2}
&&\frac{\rho^3\sin2\theta}{\sqrt{1-\rho^2\sin^2\theta}}-2N_\perp\cos2\theta\sqrt{1+\rho^2}\cos\eta+2N_\parallel\sin2\theta=0,\\
\label{Ap3}
&&\sin2\theta\sin\eta=0.
\eea
Equation (\ref{Ap3}) has two solutions:
\begin{eqnarray}
\label{5}
i)&&\theta=\frac{\pi n}{2}, \quad n\in Z\\
\label{4}
ii)&&\eta =0, \pi.
\end{eqnarray}

{\it Solution i).} The stationary points on the axis $\theta=0,\pi$ can exist provided that $\partial V/\partial\rho=0$ and $\partial V/\partial \theta=0$ for any $\psi$, which is not the case as one can see in Eqs (\ref{Ap1} -- \ref{Ap2}).
As to the equatorial plane $\theta=\displaystyle\frac{\pi }{2}$,
Eqs (\ref{Ap1} -- \ref{Ap2}) have the next solutions:
\begin{eqnarray}
\label{7}
\rho^2=\frac{N_\parallel^
{2/3}}{2^{1/3}}\left[\sqrt[3]{1+\sqrt{1+\frac{4N_\parallel^2}{27}}}+\sqrt[3]{1-\sqrt{1+\frac{4N_\parallel^2}{27}}}\right]; \label{4n}\quad
\eta=0,\;\pi.
\end{eqnarray}
Coordinate $\rho$ in Eq. (\ref{7}) increases monotone as $N_\parallel$ increases, and  asymptotically approaches the value of unity as $N_\parallel\to\infty$. For small $N_\parallel$  it takes the value $\rho\approx N_\parallel^{1/3}.$

{\it Solution ii).}
Substituting $\eta=0, \pi$
 into Eqs (\ref{Ap1}) and (\ref{Ap2}) we obtain two equations for $\rho$ and $\theta$ of the stationary points:
\bea
\label{1-1}
&&\frac{\rho^3\sin\theta}{\sqrt{1-\rho^2\sin^2\theta}}+\frac{\varepsilon N_\perp\cos\theta}{\sqrt{1+\rho^2}} -N_\parallel\sin\theta=0,\\
\label{2-1}
&&\frac{\rho^3\sin2\theta}{\sqrt{1-\rho^2\sin^2\theta}}-2\varepsilon N_\perp\cos2\theta\sqrt{1+\rho^2}
+2N_\parallel\sin2\theta=0,
\eea
where $\varepsilon=1$ for $\eta= 0$ and $\varepsilon=- 1$ for $\eta= \pi$. Solution of these equations  gives the lines at which the stationary points are lying
\begin{eqnarray}\label{line}
\tg\theta=-\varepsilon\frac{3\cot\alpha+q\sqrt{9\cot^2\alpha+8+4\rho^2}}{2\sqrt{1+\rho^2}},
\end{eqnarray}
and equation for coordinate $\rho$ at these lines:
\be
\rho^6Q^2[4+4\rho^2+Q^2]^2-N^2\sin^2\alpha[4+4\rho^2+(1-\rho^2)Q^2][2+Q\cot\alpha]^2=0,
\ee
where $Q=3\cot\alpha+q\sqrt{9\cot^2\alpha+8+4\rho^2}$ and $q=\pm 1$ is the sign of the particle charge. If $N\ll1$ and $\rho\ll1$  these equations transform to Eqs (36) and (37) of Paper I. The lines given by Eq. (\ref{line}) are plotted in Fig. \ref{fig7} for $\alpha=\pi/3$.
%%%%%%%%%%%%%%%%%%%%%%%%%%%

\end{document}